\documentclass[fleqn,usenatbib]{mnras}

\usepackage{newtxtext,newtxmath,soul}
\usepackage{rotating}
\usepackage[T1]{fontenc}
\usepackage{ulem}

\DeclareRobustCommand{\VAN}[3]{#2}
\let\VANthebibliography\thebibliography
\def\thebibliography{\DeclareRobustCommand{\VAN}[3]{##3}\VANthebibliography}

\usepackage{graphicx}	% Including figure files
\usepackage{amsmath}	% Advanced maths commands
\usepackage{adjustbox}
\usepackage{subcaption}

\title[V407 Lup]{V407 Lup, an intermediate polar nova}

\author[Orio et al.]{M. Orio$^{1,2}$\thanks{E-mail: orio@astro.wisc.edu},
M. Melicher\v{c}\'ik$^3$,
S. Ciroi$^4$,
V. Canton$^4$,
E. Aydi$^5$,
D. A. H. Buckley$^{6,7,8}$,
A. Dobrotka$^{3}$ ,
\newauthor
G.J.M. Luna$^{9}$, and
J. Ness$^{10}$
\\
$^{1}$INAF-Osservatorio Astronomico di Padova, vicolo Osservatorio, 5,
35122 Padova, Italy \\
$^{2}$ Department of Astronomy, University of Wisconsin, 475 N. Charter Str., Madison WI 53706, WI, USA \\
$^{3}$Advanced Technologies Research Institute, Faculty of Materials Science and Technology in Trnava, Slovak University of Technology in Bratislava,\\
Bottova 25, 917 24 Trnava, Slovakia\\
$^{4}$Department of Physics and Astronomy, Universit\`a di Padova, vicolo dell’Osservatorio 3, I-35122 Padova, Italy\\
$^{5}$Department of Physics and Astronomy, Center for Data Intensive and Time Domain Astronomy, Michigan State University, East Lansing, MI 48824, USA \\
$^6$South African Astronomical Observatory, P.O Box 9, Observatory, 7935 Cape Town, South Africa\\
$^7$Department of Astronomy,
University of Cape Town, Private Bag X3, Rondebosch 7701, South Africa \\
$^8$Department of Physics, University of the Free State, PO Box 339, Bloemfonein 9300, South Africa\\
$^{9}$CONICET-Universidad Nacional de Hurlingham, Av. Gdor. Vergara 2222, Villa Tesei, Buenos Aires,
Argentina \\
$^{10}$XMM-Newton Science Operations Centre, European Space Astronomy Centre, Camino Bajo del Castillo s/n, Urb. Villafranca del Castillo, 28692,\\
 Villanueva de la Cañada, Madrid, Spain)
}

\date{Accepted XXX. Received YYY; in original form ZZZ}

\pubyear{20xx}

\begin{document}
\label{firstpage}
\pagerange{\pageref{firstpage}--\pageref{lastpage}}
\maketitle
\begin{enumerate}
    \item 
\end{enumerate}
% Abstract of the paper
\begin{abstract}
We  present X-ray and optical observations of nova V407 Lup (Nova Lup 2016), previously well monitored in outburst, as it returned to quiescent accretion. The X-ray light curve in 2020 February revealed a clear flux modulation with a stable period of 564.64$\pm$0.64 s, corresponding to the period measured in outburst and attributed to the spin of a magnetized white dwarf in an intermediate polar (IP) system. This detection in quiescence is consistent with the IP classification proposed after the nova eruption. The {\sl XMM-Newton} EPIC X-ray flux is is $\simeq 1.3 \times 10^{-12}$ erg cm$^{-2}$ s$^{-1}$ at a distance, most likely,
 larger than 5 kpc, emitted in the whole 0.2-12 keV range without a significant cut-off energy.
 The X-ray spectra are complex; they can be fitted including a power law component with a relatively flat slope (a power law index $\simeq$1), although, alternatively, a hard thermal component at kT$\geq$19 keV also yields a good fit. The SALT optical spectra obtained in 2019 March and 2022 May are quite typical of IPs, with strong emission lines, including some due to
 a high ionization potential, like He II at 4685.7 \AA. Nebular lines of O [III] were prominent in 2019 March, but their intensity and equivalent width appeared to be decreasing during that month, and they were no longer detectable in 2022, indicating that the nova ejecta dispersed. Complex profiles of the He II lines of V407 Lup are also characteristic of IPs, giving
 further evidence for this classification.
\end{abstract}

% Select between one and six entries from the list of approved keywords.
% Don't make up new ones.
%%%%%%%%%%%%%%%%% BODY OF PAPER %%%%%%%%%%%%%%%%%%
\section{Introduction}
In this paper we present observations, done in quiescence a few years after a nova outburst, of  V407 Lup, a nova whose outburst is thought to have occurred a highly magnetized white dwarf (WD) \citep{Aydi2018}. 

V407 Lup has been classified as an {\it intermediate polar} (IP), a subclass of cataclysmic variables (CVs), interacting binary systems in which a WD accretes matter from a non-degenerate stellar companion \citep[see][]{Ferrario2015}. The separation of the two stars is usually less than a few solar radii, so the interaction between the components occurs via Roche lobe overflow.
 CV systems with  strongly magnetic WDs are
classified as either polars and intermediate polars (IPs). 
Polars contain WDs with a magnetic field strong enough ($> 10^{6} G$) 
to completely prevent the formation of an accretion disk, and to
 cause synchronization
of the spin period of the WD (P$_{\rm spin}$) with the orbital period of the
 binary (P$_{\rm orb}$). 

\newpage
 
In IPs, the magnetic field is less strong ($10^{5}-10^{6}$ G), 
allowing the formation of a disk (or
 in some cases, rather a ``ring'')
 that is disrupted near the Alfv\`en radius.
The flow in IPs, close to the WD, 
consists of two magnetically confined accretion
 ``curtains'' fed by the disk or ring \citep{Ferrario1993, Hellier1995}. 
 The IPs  
 do not have a synchronised P$_{\rm spin}$ and P$_{\rm orb}$ \citep[for a review see][]{Ferrario2015} and the  two periods in IPs have typical values that range 
from 1.6 h to 13 h for the orbital and from 0.5 to 70 minutes for the rotational 
 one \citep{warner1995}.
 
 Where the accretion stream impacts the WD surface a strong shock is formed,
 and since the matter is practically in free fall and its
 velocity depends only on the gravitational well of the WD, {\it
 the temperature of
 the plasma is  directly proportional to the WD mass}.  The
 temperature is such that thermal emission ensues, with peak in the  
 X-ray range; the spectrum can be fitted with
 either photoionization, or assuming a shocked  plasma
 in collisional ionization equilibrium. In some cases 
 high resolution X-ray spectra are best fitted by modeling a stationary 
 cooling flow that assumes a steady state isobaric radiative cooling
 \citep[see][]{Mukai2003, Ferrario2015}.
 {\it The X-ray luminosity is proportional to the mass accretion
 rate}, which is a second reason that
 makes X-ray observations a powerful diagnostic of the system's
 parameters and physics. Because the WDs of IPs are  generally inclined 
with respect to their
rotation axes, the asynchronous primary is an oblique rotator.
 When the accretion curtain is along the line of sight,
 it causes additional absorption of the X-ray flux emitted near the pole
 at a given phase as the WD rotates.  
 The amplitude and characteristics of the modulation
 depend on the shock height and angle subtended by the ``curtains''.
The X-ray flux modulation with the WD spin period is one
of the main observational properties of IPs, and it is
 the best indirect evidence  of the IP nature when direct measurements
 of circular polarization are not possible because the system is 
 at large distance and not sufficiently luminous.

 Until about 20 years ago, only a handful of magnetic CVs where
 observed in a nova outburst, but more novae, in
 addition to V407 Lup, have been identified 
 as belonging to this class in recent years: V2487 Oph
 \citet{Hernanz2002},  V2491 Cyg \citep{Zemko2015},
 V4743 Sgr \citep{Zemko2016, Zemko2018}, and V1674 Her \citep{Drake2021}.   

The root cause of a nova outburst is a thermonuclear runaway (TNR). As the accreted envelope accumulates on the WD surface, temperature and density rise at the base of the envelope, leading to nuclear burning in a shell. Once the accreted layer reaches a critical mass, depending on the WD mass and accretion rate, nuclear burning in most cases becomes unstable and a  TNR follows, leading to a nova explosion \citep{Yaron2005, Starrfield2012, Wolf2013}. 
 The fact that the mass accretion rate and WD mass are fundamental
 parameters determining the outcome of the outburst and the system
 evolution makes X-ray observations of the systems, as they return
 to quiescence, particularly interesting, especially for IPs.  
After the TNR, an external shell of gas on the WD surface is expelled
 in a very fast wind caused by radiation pressure and/or filling
 of both Roche lobes in a common envelope \citep[see][]{Wolf2013, Shen2022},
  expanding into the interstellar medium (ISM) at thousands 
 kilometer per seconds \citep[see, among others, for both theory and observations][]{Yaron2005, Aydi2020}.

The nova outburst of V407 Lup  was discovered in 2016 September by the All-Sky Automated Survey for Supernovae (ASAS-SN) on HJD 2457655.5 at $V = 9.1$ \citep{stanek}. The nova was classified as very fast, with a decline time by 2 magnitudes $t_{2} \la 2.9$ $d$ \citep{Aydi2018}. V407 Lup reached the optical peak after 1.4 days from the time of discovery and went through the three main phases, namely early He/N, FeII and late He/N, during the firsts three weeks (in the context of the spectroscopic evolution presented in \citealt{Aydi2024}). By day 19, strong forbidden lines of O and N were detected, indicating that the nova entered the nebular phase \citep{Aydi2018}. 
Timing analysis of the X-rays in the supersoft X-ray phase showed two periodicities of 3.57h and 565s, which were attributed  respectively the $P_{orb}$ and the $P_{spin}$ of the WD, suggesting that the system is an IP \citep{Aydi2018}, even if the X-ray modulation was not simply due to accretion (we will discuss this point more in detail below).\\
The distance to the nova has not been determined with precision. The GAIA parallax measured for the only possible progenitor candidate  within 1 arcsec from the nova position is consistent with a geometric distance of 4.7$^{+6.5}_{-2.5}$ kpc and a photogeometric distance of 8.6$^{+4.4}_{-2.6}$ kpc \citep{Bailer2021}. Assuming a distance of 10 kpc, in outburst the peak supersoft X-ray luminosity was consistent with the Eddington luminosity, close to 10$^{38}$ erg s$^{-1}$  for a 1 M$_\odot$ star \citep{Aydi2018}.

\section{X-ray Observations}
V407 Lup was observed with the European Photon Imaging Camera (EPIC) of the \textit{XMM-Newton} on 2020 February 26. We focus especially on the data of the EPIC camera, which has a pn CCD array \citep{struder} and two Metal Oxide Semi-conductor CCD arrays (MOS1 and MOS2) \citep{turner}). The instruments were operated in full window and imaging mode. Observations started on 2020-02-26 at UT 02:57 and the exposure time was 48 ks. The \textit{XMM-Newton} data were reduced and analyzed using the Science Analysis System (SAS) software. As we describe in Subsection 2.2, we also examined the high resolution spectrum with the Reflection Grating Spectrometers (RGS) in the 5-35 \AA \ (0.35-2.48 keV). Although the signal-to-noise in the RGS was very poor, a few prominent
emission lines were detected, completing the picture offered by the EPIC instruments.
A count rate of 0.151$\pm$0.030 cts s$^{-1}$ was measured with the EPIC pn in the 0.2-12 keV range. The MOS1 and MOS2 count rates in the
 0.3-12 keV range were 0.0394$\pm$0.0013 cts s$^{-1}$ and 0.0476$\pm$0.0014 cts s$^{-1}$, respectively. The count rate measured with the RGS co-adding the RGS1 and RGS2 exposures was 0.0080$\pm$0.0008 cts s$^{-1}$ in the RGS range.

\subsection{Spectral Analysis: EPIC}
\begin{figure*}
\includegraphics[height=6cm]{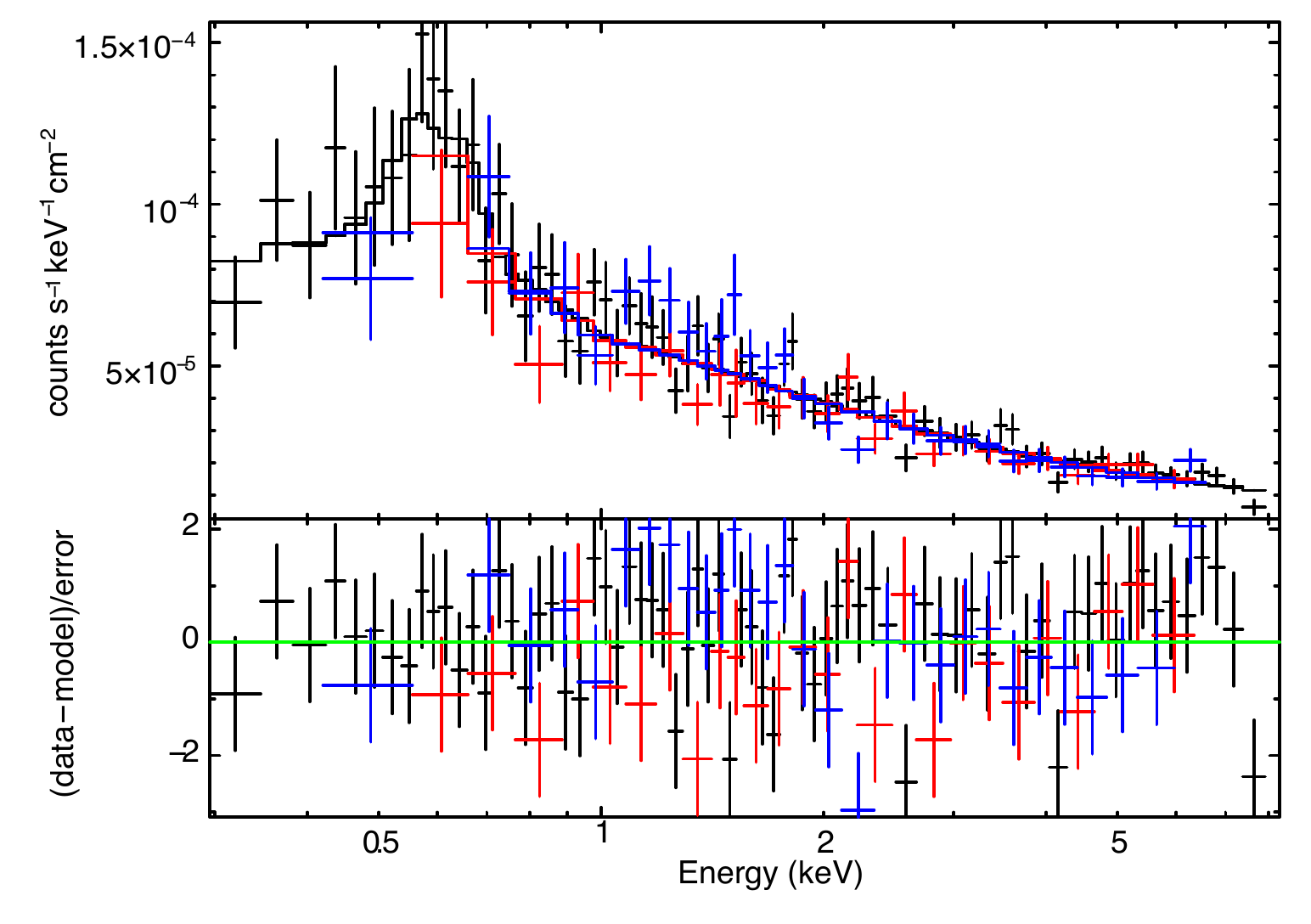}
\includegraphics[height=6cm]{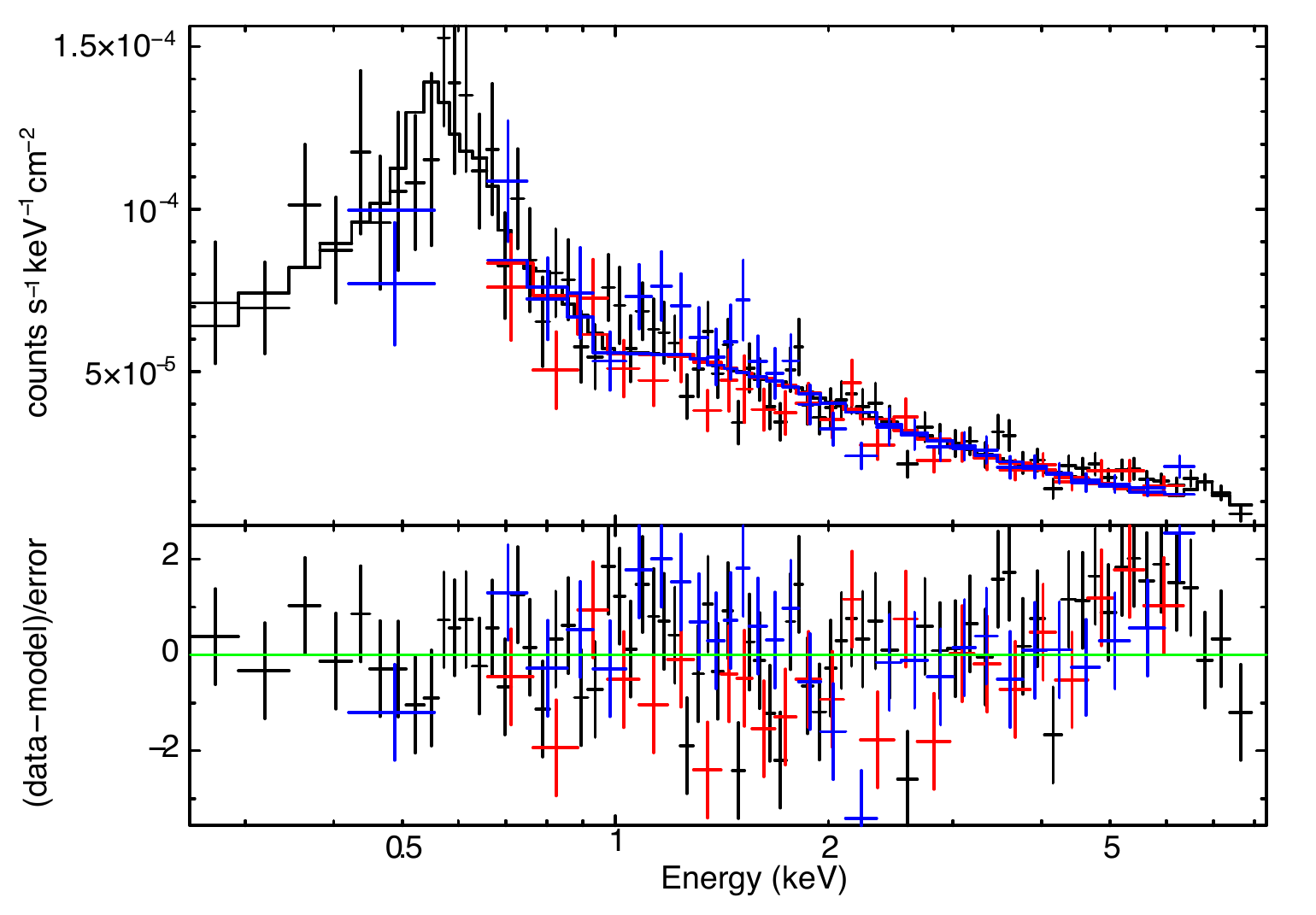}
\caption{The spectra measured with EPIC pn (in black), EPIC-MOS1 (in red) and EPIC-MOS2 (in blue). On the left, a fit with a power law with index
$\simeq$1 and a VAPEC thermal component at 0.19 keV temperature (Model 2)
and on the right, a fit with  a thermal component (VAPEC, or collisional ionization in equilibrium) at 0.15 keV,
 and a second hard thermal component at $\simeq$68 keV ( Model 3). \label{fig:1}}
\end{figure*}

Spectroscopic analysis of the three XMM-Newton EPIC spectra (pn, MOS1 and MOS2 cameras) was performed using the XSPEC v12.12.1 fitting engine.  We did
 not find a good fit with a cooling flow model and we could not measure
 spectra lines that would prove photoionization. However,
 we obtained reasonable fits with   three different composite models, all involving  a thermal plasma in collisional
ionization equilibrium, and both
 a very soft and a very hard component. In all models we assumed N(H)$\geq 1.4 \times 10^{21}$ cm$^{-2}$,  consistent with the value indicated by the HEASOFT
tool nH (\url{https://heasarc.gsfc.nasa.gov/cgi-bin/Tools/w3nh/w3nh.pl}),
  although there can be a small deviation from
 the average value in the direction of the nova, and/or some intrinsic absorption in the system. Fig. \ref{fig:1} presents the background-subtracted spectra of V407Lup, extracted in the 0.3-10 keV range of the two MOS and in the 0.2-10 keV range of the pn, fitted using our models Model 2 and Model 3. It is evident that there is very little difference between the two fits.

{\bf Model 1} is TBABS x (blackbody+PCFABS$\times$APEC), where APEC is a model of thermal plasma in collisional ionization equilibrium with solar abundances \citep{apec}, following examples of other
novae that are also IPs \citep{Zemko2015, Zemko2016, Sun2020}. TBABS is the T\"ubingen-Boulder ISM absorption model in XSPEC \citep{Wilms2000}. 
Since we could not obtain a good fit with only the TBABS model, we added in the XSPEC fit 
the partially covering absorber PCFABS (\url{https://heasarc.gsfc.nasa.gov/xanadu/xspec/manual/node258.html}) which has been often used to describe well the absorption due to the accretion column near the poles, assuming the IP nature of the WD \citep[see review by][]{Mukai2017}. 
The best fit value return of 2.5 $\times 10^{21}$ $cm^{-2}$,  possibly implies some intrinsic absorption in the whole system. The PCFABS column density turned out to be very high, namely  $N(H) = 11.45 \times 10^{22}$ $cm^{-2}$, covering 44\% of the X-ray emitting source, and this value is larger than found in other IPs. The other parameters are shown in the first column of Table \ref{tab:1}. In this model, almost all the unabsorbed flux is due to the APEC component, which converges only with the highest temperature calculated in the model, 64 keV. The fit  yields a value of $\chi^2$/(degrees of freedom)$\simeq$1.2, but there is some unexplained soft excess in the softest pn portion of the spectrum. We experimented by fixing the temperature to a lower value, and obtained $\chi^2$/(degrees of freedom)$\geq$1.48 with kT = 19.62 keV. Thus we concluded that the APEC temperature is not really constrained by our data. We note that the blackbody flux resulting from the fit is small, and for any distance between $\simeq$2.5 and $\simeq$ 10 kpc it is consistent with an origin in a small region (presumably at the polar caps),  not from the entire WD surface.

\begin{figure*}
    \centering
\includegraphics[width=17.5cm]{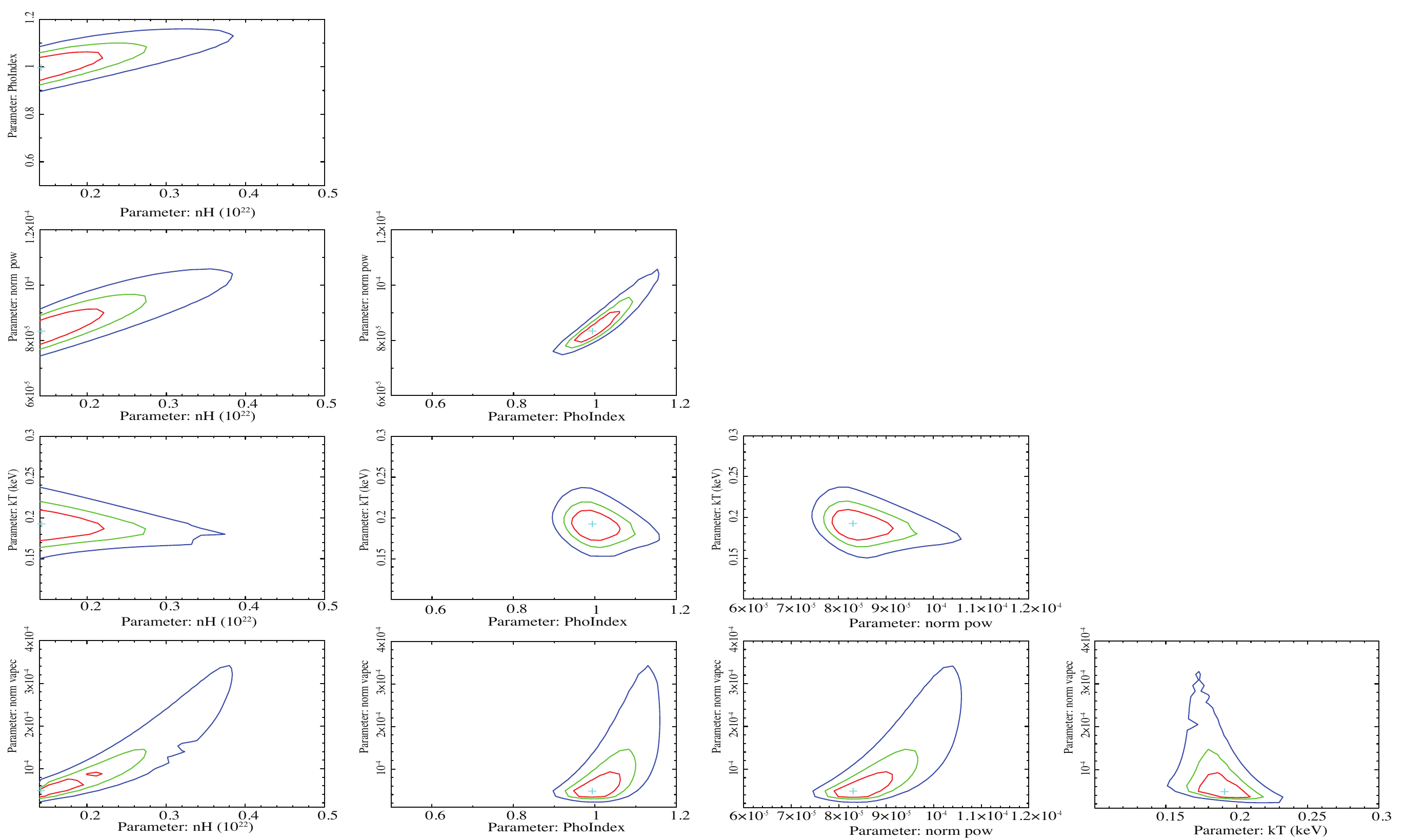}
    \caption{Contours plots of the TBabs x(powerlaw + vapec) model parameters (N(H), kT, and normalization constants of the two models), assuming that N(H)$\geq 1.4 \times 10^{21}$ cm$^{-2}$ (see text). The red, green and blue solid lines 
    correspond respectively to the 68\%, 90\% and 99\% confidence levels.}
    \label{fig:2} 
\end{figure*}

{\bf Model 2.} Given the presence of a hard component, we also fitted the spectrum
 with a flat powerlaw and a VAPEC thermal component (APEC with varying abundances) for the soft portion of the spectrum, TBABS $\times$ (powerlaw+VAPEC). As we discuss more in detail below, a power law component is  necessary in fitting some X-ray spectra of IPs, and it may indicate a non-thermal component. This composite model does not require a partially covering absorber. The best fit parameters  are shown in Fig. \ref{fig:1} and in Table 1. We derived a photon index $\alpha \simeq 1$, and the
flux is mainly due to this powerlaw component. We varied the C, N, Ne, Mg and Fe abundances (which may be non-solar in novae) between  1/10th and 10 times the solar abundances .  For the thermal component we obtained $kT=0.19$ keV in the fit to all instruments, both fitted separately and together.
Fig. \ref{fig:2} shows a ``corner plot'', namely the 1,2, and 3 $\sigma$ confidence contours of the acceptable values obtained by varying only two at once among the following parameters: N(H), photon index $\alpha$, normalization of both components and VAPEC temperature with the {\it steppar} XSPEC command. The Figure indicates that these four parameters are quite well constrained in a narrow range. Allowing the abundances of several elements to be free parameters improves the fit. We obtained enhanced carbon, about solar nitrogen and carbon and largely depleted other elements, but the abundance determination has large errors,
because the EPIC cameras do not resolve individual lines.so these parameters are not shown in the corner plot. 

{\bf Model3.} Finally, we experimented with two VAPEC components to fit both the hard and soft portion of the spectrum. We did not need
 PCFABS for the fit, but initially we used variable abundances as free parameters. 
The best fit was still obtained with kT$\simeq$69 keV (maximum temperature in the XSPEC VAPEC model)
  for one of the components, and with kT=0.16 keV for the soft component. We allowed all the abundance to vary again between
 1/10th and 10 times solar, and the abundances of several elements resulted depleted in the best fit, except for solar oxygen and 
enhanced carbon, calcium, argon, and iron. However, again
the errors on the abundances are very large; the abundances thus are not well constrained by the available data.
The best fit was obtained with
a blue shift velocity as high as 6000 km s$^{-1}$, although the value
of $\chi^2$/d.o.f. increases only slightly assuming that the velocity is even equal to 0. 
\begin{table*}
\caption{Best fit parameters of Model 2 and Model 3 for the three
 EPIC instruments together}. 
\begin{center}
\begin{tabular}{l c c c c } 
 \hline
 Parameter &  Model 1 & Model 2  & Model 3\\ 
 \hline
N(H)\textsuperscript{a} $\times 10^{22}$ cm$^{-2}$ &  0.25$\pm0.1$ & 0.14$_{-0.14}^{+0.08}$ & 3.57$_{0.69}^{+0.78}$ \\
  & & & \\
N(H)\textsubscript{pc} $ \times 10^{22}$ cm$^{-2}$  & 11.45$_{-4.63}^{+9.36}$ & -- & -- \\
& & & \\
 CvrFract & 0.45$_{-0.09}^{+0.07}$ & -- & -- \\
 & & & \\
PhoIndex($\alpha$)\textsuperscript{b} & -- &  0.99$_{-0.05}^{+0.07}$ & --  \\
& & & \\
Norm\textsubscript{pow}$\times 10^{-5}$ & -- & 8.32$_{-0.45}^{+0.88}$ & --  \\
& & & \\
T\textsubscript{bb}(eV) & 90$_{-8}^{+11}$ & -- & --  \\
& & & \\
Norm\textsubscript{bb} & 7.21$_{-5.3}^{+11.4} \times 10^{-6}$ & -- & -- & \\
& & & \\
T\textsubscript{apec}(keV) & 19.62$_{-7.19}^{+21.36}$ & -- & -- \\
& & & \\
Norm\textsubscript{apec}$\times 10^{-4}$ & 8.8$_{-0.9}^{+1.31}$ &  -- &  --\\
& & & \\
T\textsubscript{vapec1}(keV) & -- &  0.19$_{-0.04}^{+0.08}$ & 0.15$\pm0.01$ \\
& & & \\
Norm\textsubscript{vapec1}$\times 10^{-5}$ & -- &  4.12$_{-1.33}^{+5.09}$ &  \\
& & & \\
T\textsubscript{vapec2}(keV) & --  & -- &  68.5$_{-19.5}$ \\
& & & \\
Norm\textsubscript{vapec2}$ \times 10^{-4}$ & -- & -- & 4.63$_{-2.87}^{+13.06}$ \\
& & & \\%
C &  -- & 8.23$_{-8.23}^{+25.27}$ & 10.0 (u.l.) \\
& & & \\
N & -- & 1.15$_{-1.16}^{+7.48}$ & 0.1 \\
 & & & \\
Ne & -- & 0.52$_{-0.51}^{+2.10}$ & 0.96$_{-0.96}^{+1.18}$ \\
 & & &  \\
Ar & -- & -- & 10.0 \\
 & & & \\
Ca & -- & -- & 10.0 \\
 & & & \\
Fe  & --  & 1.053$_{-1.052}^{+16.41}$ & 10.0 \\
 & & & \\
Flux$_{\rm total, abs.}\times 10^{-12}$ erg cm$^{-2}$ s$^{-1}$  & 1.21 & 1.25 & 1.36 \\
 & & & \\
Flux$_{\rm total, unabs} \times 10^{-12}$ erg cm$^{-2}$ s$^{-1}$ & 2.33  & 1.45 & 3.67 \\
 & & & \\
Flux$_{\rm pow} \times 10^{-12}$ erg cm$^{-2}$ s$^{-1}$  &  -- & 1.22 & --  \\
 & & & \\
Flux$_{\rm pow, unabs} \times 10^{-12}$ erg cm$^{-2}$ s$^{-1}$  &  -- & 1.32 & -- \\
 & & & \\
Flux$_{\rm bb}  \times 10^{-12}$ erg cm$^{-2} $ s$^{-1}$                & 0.03 & --  & -- \\
 & & & \\
Flux$_{\rm bb,unabs}  \times 10^{-12}$ erg cm$^{-2} $ s$^{-1}$          & 0.47  & -- & -- \\
 & & & \\
Flux$_{\rm apec}  \times 10^{-12}$ erg cm$^{-2}$ s$^{-1}$ &
 1.54 & -- & -- \\
 & & & \\
Flux$_{\rm apec, unabs}  \times 10^{-12}$ erg cm$^{-2} $ s$^{-1}$ & 1.86 & -- & -- \\
 & & & \\
Flux$_{\rm vapec1}  \times 10^{-12}$ erg cm$^{-2} $ s$^{-1}$ &
 -- & 0.03 & 4.67 \\
 & & & \\
Flux$_{\rm vapec1,unabs}  \times 10^{-12}$ erg cm$^{-2} $ s$^{-1}$ 
 -- & -- & 0.13 \\
 & & & \\
 Flux$_{\rm vapec2}  \times 10^{-12}$     erg cm$^{-2}$ s$^{-1}$  & --  & -- & 1.31 \\
  & & & \\
Flux$_{\rm vapec2,unabs}  \times 10^{-12}$ erg cm$^{-2}$ s$^{-1}$  & --  &  -- &  1.53 \\
 & & & \\
 $\chi^2$/d.o.f. & 1.48  & 1.51 & 1.46  \\
 \hline
\end{tabular}
\end{center}
 \label{tab:1}
 \end{table*}
\begin{figure}
\resizebox{\hsize}{!}{\includegraphics[angle=0]{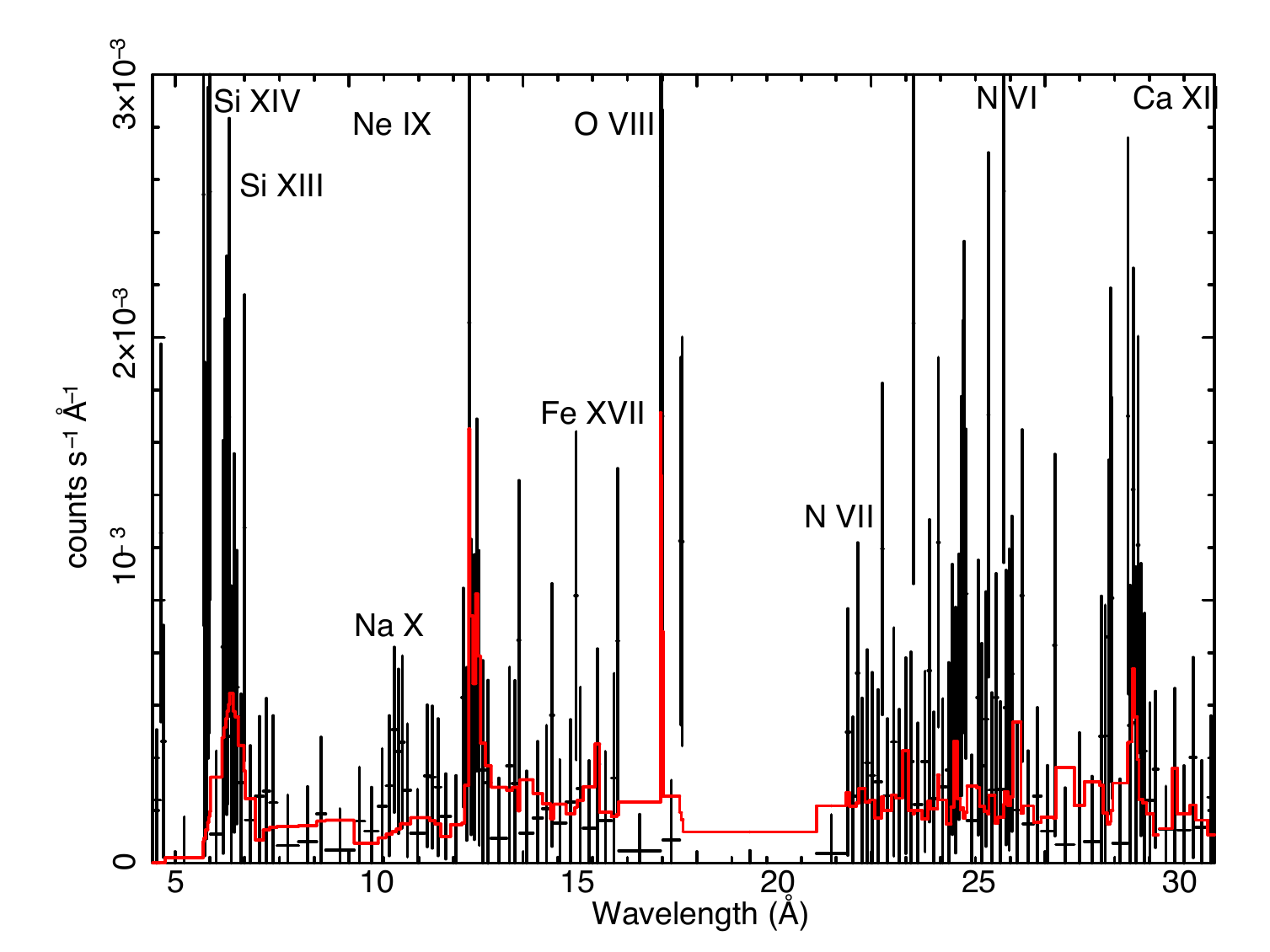}}
\caption{The coadded RGS1 and RGS2 spectra, and a fit with a model with a BVAPEC component at T=1.27 kev and one at  T=0.22 keV.}
\end{figure}
\subsection{The RGS high resolution spectrum.}
  The coadded spectrum of both RGS1 and RGS2 spectrometers is shown in Fig. 3. An approximate fit with two components
 modeled with BVAPEC (a model in which we added line broadening to VAPEC)
 is also shown in the Figure, with temperatures of 0.25 and 1.3 keV. The RGS are not sensitive to the ``hard'' portion of the spectrum, but they do show that,  if the emission lines are attributed to collisional ionization equilibrium thermal components, they originate in regions with more than one  temperature even in the lower energy range. Thus, there may be 3 or
   more different regions with different temperature to explain the spectrum across the whole 0.2-13 keV range.
    BVAPEC includes only solar abundance of sodium, so the Na line could not be fitted. The other elements were allowed to vary; the fit converged with silicon, sulphur, aluminium, argon and calcium enhanced by a factor of 10; carbon and neon by a factor $\simeq$1.5,
   and carbon, nitrogen and iron depleted by 60\%-80\%. However,
    the signal-to-noise is too poor to obtain a rigorous fit, and even an attempt to fit the RGS together with the EPIC cameras did not produce a 
    statistically significant fit, due to the low signal-to-noise in the RGS spectrum.
\subsection{Timing analysis}
\begin{figure}
\resizebox{\hsize}{!}{\includegraphics[angle=0]{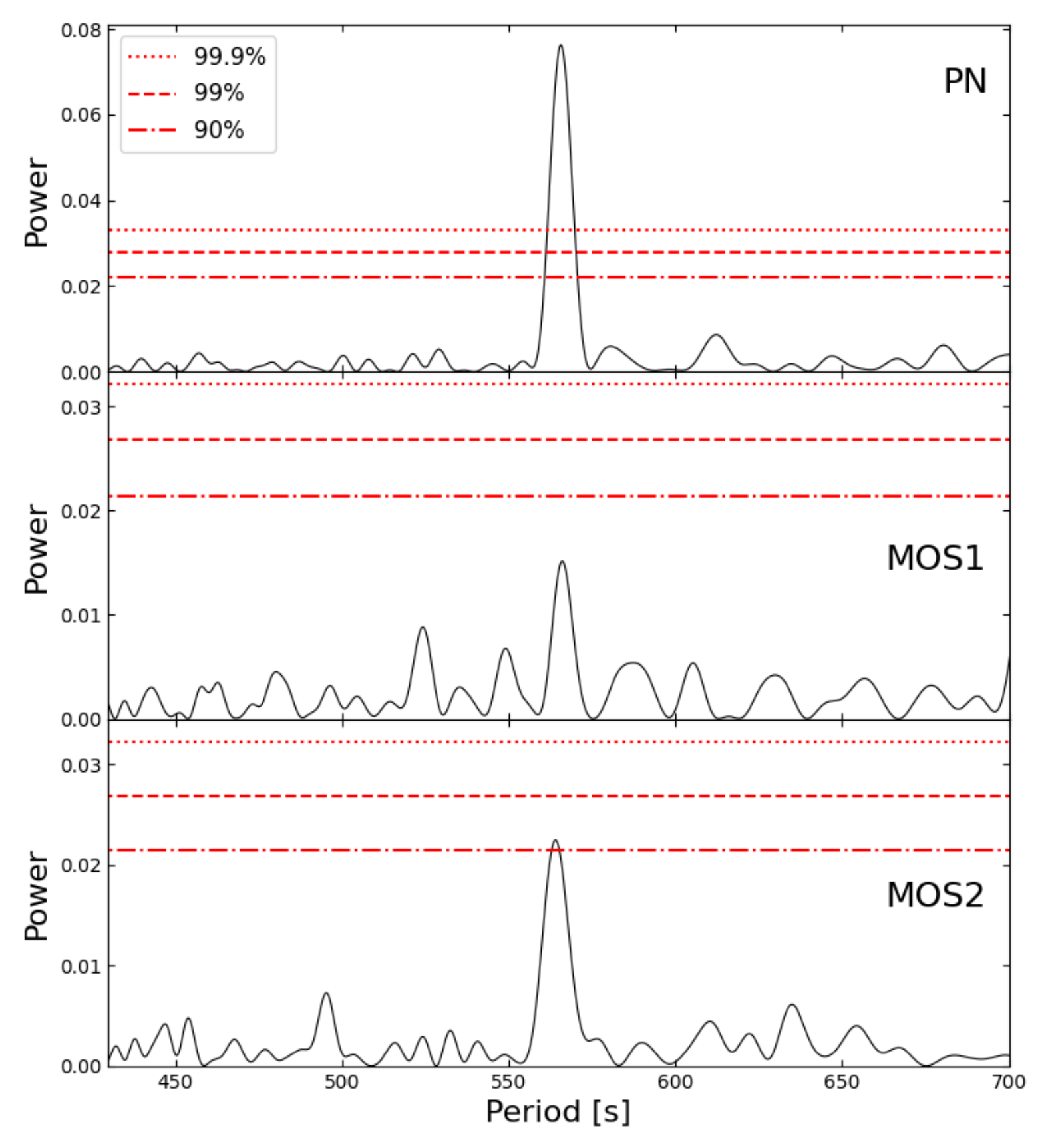}}
\caption{Periodograms calculated from all three X-ray light curves from \textit{XMM}.The power of a peak at the 90\%, 99\% and 99.9\% confidence levels is indicated by the red horizontal lines.}
\label{periodogram}
\end{figure}

For the timing analysis, the pn, MOS1, and MOS2 light curves were extracted from a circular region with a radius of 15\arcsec\ centered on the source, while the background was extracted from an offset region with the same radius. We binned the light curves with bins of 50 s for the MOS and 60 s for the pn, and excluded the last 4.5\,ks from the timing analysis for the three instruments, because of elevated background.
For this analysis, we used the Lomb-Scargle algorithm\footnote{We used 
the python's package {\tt Astropy} 
(\citealt{astropy_collaboration2013,astropy_collaboration2018,astropy_collaboration2022}).} 
\citet{scargle1982}. The three periodograms we obtained are shown in 
Fig. ~\ref{periodogram}. The confidence of the peak in pn periodogram is 
far above 99.9\%, proving very significant variability with 
periodicity of 565.64\,s. The MOS1 periodogram has a much lower confidence  of 34\%, while the MOS2 periodogram has again a high  confidence, slightly above 93\%.
In order to obtain  an estimate of the statistical error on the 565.64\,s period obtained with the pn, we fitted a Gauss function to the pn peak. The sigma parameter as the error estimate is 3.10\,s. Since this error depends on the periodogram resolution, we performed another estimate using simulations. We fitted a sine function to the pn light curve with periodicity fixed to 565.64\,s and simulated ten thousands light curves with added Poisson noise. For every simulated light curve we calculated a periodogram and we searched the peak periodicity. We fitted the resulting histogram of periodicities measured from the simulated lightcurves with a Gauss function, and obtained a value for the statistical uncertainty $\sigma=0.33$ (namely, the period with a 1 $\sigma$ error is 565.68$\pm$0.33 s). The same exercise yielded these results: 566.18$\pm$0.81 s for the MOS1 and 564.36$\pm$0.92 s for the MOS2. Thus, despite the low confidence level obtained with the MOS1 data, the three periods are consistent with each other. Given the very significant pn result, with its high confidence level, we concluded that the periodic modulation is definitely present in the X-ray light curve.
\section{The optical spectra}
Optical spectra in the 3156-6318 \AA \ range were obtained with the Robert
 Scobie Spectrograph RSS and the PG 900 grating (spectral resolution
 R$\approx$800) at the SALT 10m telescope. Fig. 5 shows the result of 4 exposures between 2019 February 22 and 2019 March 27, without significant changes.
However, Fig. 6 shows the spectrum obtained with the same telescope and instrumental setup on 2022 May 10. All the spectra are characterized by strong emission lines, not only of the Balmer series, but also lines with high ionization potential, especially He II at  4685.7 \AA,  which is due to a transition with a ionization potential of 54.4 eV. This line was detected with about the same
flux above the continuum both in 2019 February/March and 2022 May. Also the He II lines at 4541 \AA \ and 5411 \AA \ are measured at both epochs, as it is often the case of IP spectra in which the first line is detected \citep{Masetti2012, Zemko2016}. However, the nebular lines of [O III] around 5000 \AA \ were no longer detected in 2022, showing that there was nebular emission for at least 2.5 years after the nova outburst, but not any more after 5.5 years. The disappearance of [O III] and of the other nebular lines of [N II] (see Table 3), is the remarkable difference that occurred in the space of $\simeq$3 years that elapsed between the observations. 
 
Table 2 (available on line for all the measurable lines, and here included for H $\alpha$, He II and [O III]) shows the flux in the emission lines, measured above the chosen continuum, and the equivalent width. It is important to notice that, since we have no absolute flux calibration, only the level above the continuum and the equivalent width can be compared. There was already a significant decrease in equivalent width of the [O III] lines between 2019 March 16 and 2019 March 26, probably indicating that the intensity of those lines was already dropping at that epoch. 

The equivalent width of He II, H$\alpha$ and other emission lines  was not constant within the errors, and that there was no decline in intensity between 2019 and 2022. 
 However, only analysing
 the equivalent width of the lines may  not be sufficient for a physical
 picture, due to the complex structures and line profile
 evolution we see  in Fig. ~\ref{Fig:line_profiles} for He II 4686 \& 5411\,\AA, H$\beta$,
and H$\alpha$ throughout the five SALT spectral
I epochs. He II at 5411 \AA \ shows potentially consist of multiple components, 
and there are remarkable changes in the line profiles. This is expected for magnetic CVs, where different emission
line components originate in different regions of the system (e.g., accretion stream, the flow through
the magnetosphere, truncated accretion disk; \citep{Hellier_etal_1987,Rosen_etal_1987, warner1995,Schwope_etal_1997}).
The Balmer line profiles, which are still mostly dominated by emission from the nova ejecta show minimal
variability, particularly throughout the first four observations, all done within a short period  in 2019. Similarly, the He II line at 4686\,\AA \ 
shows minor changes in 2019 due to the contribution of the nova ejecta to this line and the blending with
nearby lines (the Bowen blend). We suggest that it 
will be worth, in a future project,
to monitor the flux and radial velocity of these lines over the orbital period of 3.57 h detected by \citet{Aydi2018}.

\begin{figure}
\resizebox{\hsize}{!}{\includegraphics[angle=0]{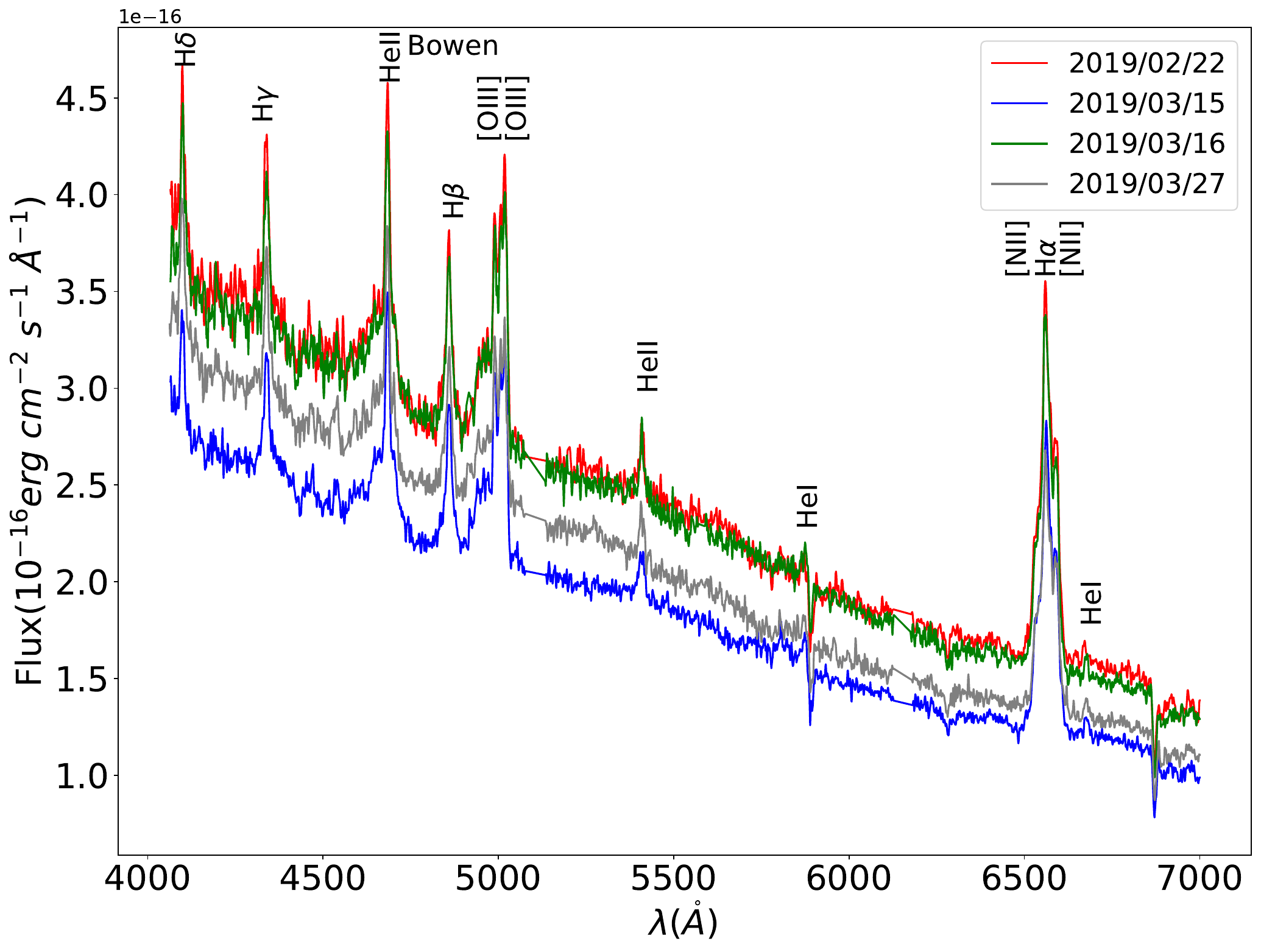}}
\caption{The (non-dereddened) optical spectra of V407 Lup taken with the SALT 10m telescope and the Robert Scobie Spectrograph (RRS) in its low resolution mode 
between 2019-02-22 and 2019-03-27. The flux scale is arbitrary and the
 spectra have been shifted for clarity.}
\label{SALT1}
\end{figure}
\begin{figure}
\resizebox{\hsize}{!}{\includegraphics[angle=0]{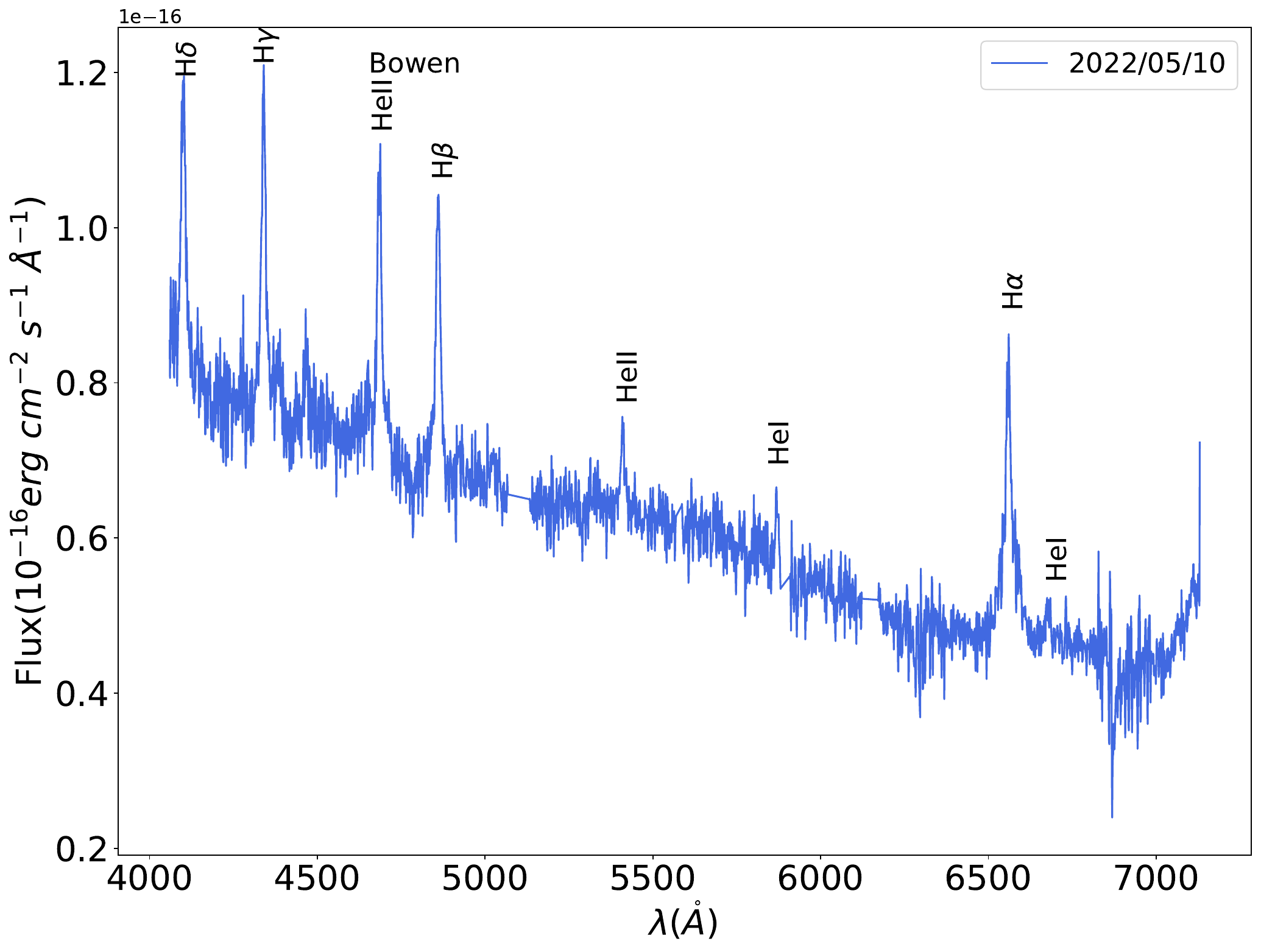}}
\caption{The (non-dereddened) optical spectrum of V407 Lup taken with the SALT 10m telescope and the RSS in its low resolution mode  on 2022-05-10.}
\label{SALT2}
\end{figure}

\begin{figure*}
\begin{center}
\includegraphics[width=\textwidth]{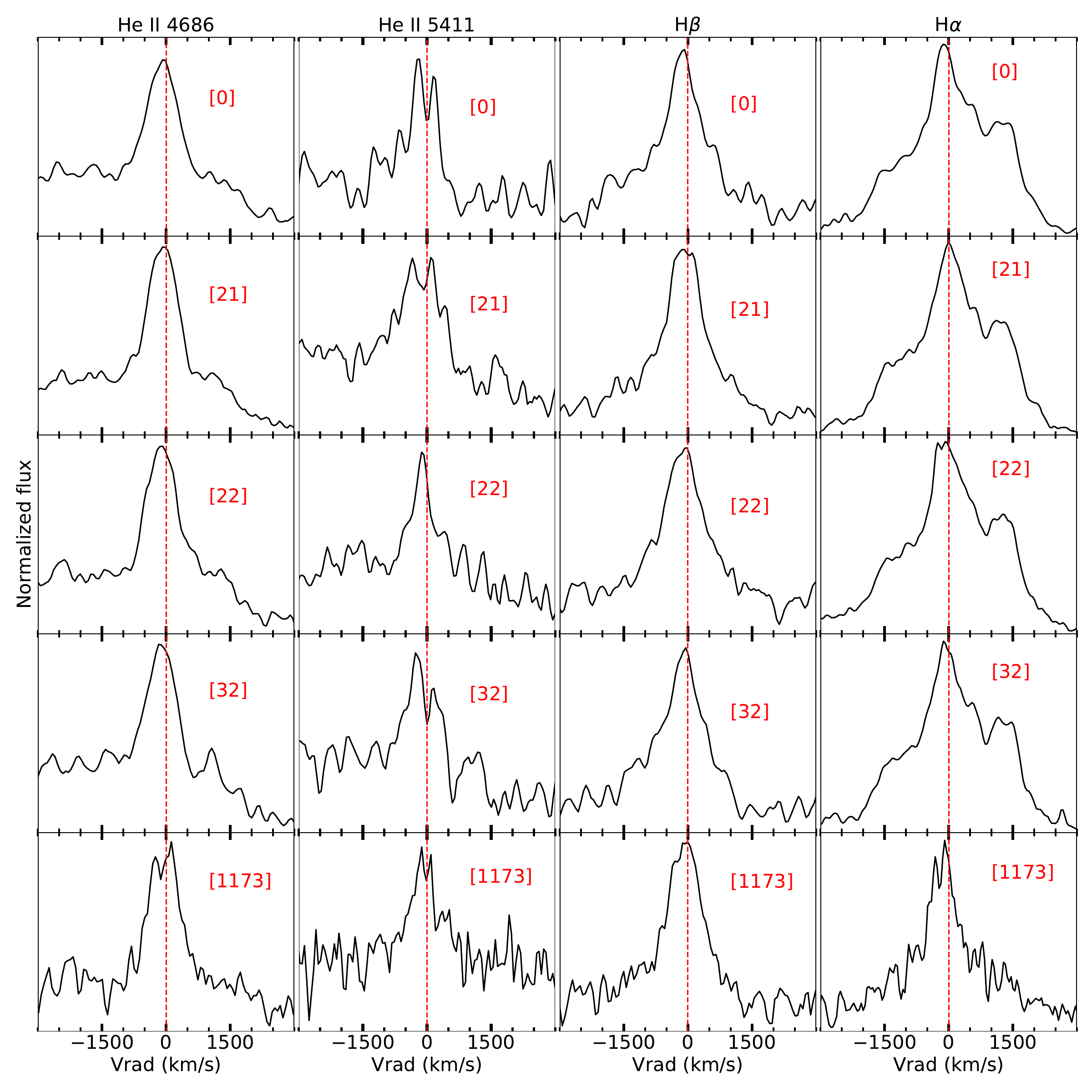}
\caption{Line profile evolution of He II 4686 \& 5411\,\AA, H$\beta$, and H$\alpha$ throughout the 5  spectral epochs of SALT observations. The red dashed lines represent $v_\textrm{rad} = 0$\,km\,s$^{-1}$. The number between brackets are days since the first spectral epoch (2019-02-22). A heliocentric correction is applied to the radial velocities.}
\label{Fig:line_profiles}
\end{center}
\end{figure*}

\begin{table*}
\caption{The nominal and measured wavelength of each line, the continuum flux level assumed under each line, the flux in the line above the assumed continuum level, the root mean square of the flux to indicate the noise level, the equivalent width with its error, the height of the Gaussian and of the  Lorentzian functions, the equivalent widths of the Gaussian and  of the Lorentzian.} 
 \begin{center}
 \rotatebox{90}{
 \begin{tabular}{l c c c c c c c c c} 
 \hline\hline
  Line & $\lambda$(\AA) & $\lambda_{\rm m}$(\AA) & Continuum & Rms\textsubscript{cont} & Line Flux & EW & Core & Gew & Lew \\
 & (\AA) & (\AA) & (erg/cm$^2$/s/\AA) & (erg/cm$^2$/s) &  (erg/cm$^2$/s/\AA) & (\AA) &(erg/cm$^2$/s/\AA) & (\AA) & (\AA)\\
 \hline
 \multicolumn{10}{c}{2019/02/22}\\
 \hline
 HeII & 4685.7 & 4684.191 & 2.956$\times 10^{-16}$ & 7.406$\times 10^{-18}$ & 2.818$\times 10^{-15}$ & 9.535$\pm$0.754 & 1.371$\times 10^{-16}$ & -- \\
 $[$O III$]$ & 4958.9 & 4955.757 & 2.740$\times 10^{-16}$ & 6.044$\times 10^{-18}$ & 1.953$\times 10^{-15}$ & 7.128$\pm$1.111 & 4.518$\times 10^{-17}$ & 40.61 & -- \\
 $[$O III$]$ & 5006.8 & 5006.914 & 2.714$\times 10^{-16}$ & 6.044$\times 10^{-18}$ & 5.631$\times 10^{-15}$ & 20.75$\pm$1.42 & 1.303$\times 10^{-17}$ & 40.61 & --\\
 H$\alpha$  & 6562.8 & 6560.505 & 1.603$\times 10^{-16}$ & 3.322$\times 10^{-18}$  & 6.124$\times 10^{-15}$ & 38.22$\pm$1.48 & 1.856$\times 10^{-16}$ & -- & 21.01 \\
 \hline
 \multicolumn{10}{c}{2019/03/15}\\
 \hline
 HeII & 4685.7 & 4684.316 & 2.282$\times 10^{-16}$ & 2.667$\times 10^{-18}$ & 2.236$\times 10^{-15}$ & 9.797$\pm$0.364 & 1.046$\times 10^{-16}$ & -- & 13.61 \\
 $[$O III$]$ & 4958.9 & 4950.532 & 2.122$\times 10^{-16}$ & 2.667$\times 10^{-18}$ & 1.795$\times 10^{-15}$ & 8.459$\pm$0.680 & 3.932$\times 10^{-17}$ & 42.89 & --\\
 $[$O III$]$ & 5006.8 & 5006.359 & 2.095$\times 10^{-16}$ & 2.667$\times 10^{-18}$ & 4.709$\times 10^{-15}$ & 22.48$\pm$0.87 & 1.031$\times 10^{-16}$ & 42.89 & -- \\
 H$\alpha$  & 6562.8 & 6562.778 & 1.230$\times 10^{-16}$ & 2.708$\times 10^{-18}$ & 5.983$\times 10^{-15}$ & 48.65$\pm$1.90 & 1.582$\times 10^{-16}$ & -- & 24.08 \\
 \hline
 \multicolumn{10}{c}{2019/03/16}\\
 \hline
 HeII & 4685.7 & 4684.016 & 2.945$\times 10^{-16}$ & 6.601$\times 10^{-18}$ & 1.390$\times 10^{-15}$ & 4.718$\pm$0.420 & 9.900$\times 10^{-17}$ & 13.19 & --\\
 $[$O III$]$ & 4958.9 & 4950.532 & 2.122$\times 10^{-16}$ & 2.667$\times 10^{-18}$ & 1.795$\times 10^{-15}$ & 8.459$\pm$0.680 & 3.932$\times 10^{-17}$ & 42.89 & --\\
 $[$O III$]$ & 5006.8 & 5006.359 & 2.095$\times 10^{-16}$ & 2.667$\times 10^{-18}$ & 4.709$\times 10^{-15}$ & 22.48$\pm$0.87 & 1.031$\times 10^{-16}$ & 42.89 & --\\
 H$\alpha$  & 6562.8 & 6562.778 & 1.230$\times 10^{-16}$ & 2.708$\times 10^{-18}$ & 5.983$\times 10^{-15}$ & 48.65$\pm$1.90 & 1.582$\times 10^{-16}$ & -- & 24.08 \\
 \hline
 \multicolumn{10}{c}{2019/03/27}\\
 \hline
 HeII & 4685.7 & 4683.656 & 2.605$\times 10^{-16}$ & 2.999$\times 10^{-18}$ & 1.205$\times 10^{-15}$ & 4.624$\pm$0.204 & 9.228$\times 10^{-17}$ & 12.26 & -- \\
 $[$O III$]$ & 4958.9 & 4951.589 & 2.436$\times 10^{-16}$ & 2.999$\times 10^{-18}$  & 1.309$\times 10^{-15}$ & 5.371$\pm$0.638 & 2.819$\times 10^{-17}$ & 43.61 & -- \\
 $[$O III$]$ & 5006.8 & 5005.869 & 2.394$\times 10^{-16}$ & 2.999$\times 10^{-18}$ & 3.693$\times 10^{-15}$ & 15.43$\pm$0.78 & 7.955$\times 10^{-17}$ & 43.61 & --\\
 H$\alpha$ & 6562.8 & 6561.745 & 1.345$\times 10^{-16}$ & 2.114$\times 10^{-18}$ & 3.055$\times 10^{-15}$ & 22.7$\pm$0.7 & 1.224$\times 10^{-16}$ & 23.45 & --\\
 \hline
 \multicolumn{10}{c}{2022/05/10}\\
 \hline
 HeII & 4685.7 & 4686.365 & 6.932$\times 10^{-17}$ & 2.272$\times 10^{-18}$ & 6.108$\times 10^{-16}$ & 8.811$\pm$0.851 & 3.558$\times 10^{-17}$ & 16.13 & --\\
 H$\alpha$ & 6562.8 & 6560.052 & 4.716$\times 10^{-17}$ & 1.531$\times 10^{-18}$ & 1.335$\times 10^{-15}$ & 28.3$\pm$2.248 & 3.260$\times 10^{-17}$ & -- & 26.07 \\
 \hline\hline
 \end{tabular}}
  \end{center}
   \end{table*}

\section{Discussion}
With the X-ray observations done once quiescent accretion had resumed,
we can confirm that the spectral and timing characteristics of V407 Lup are quite typical of an intermediate polar. The X-ray luminosity was higher than that of most CVs in quiescence. The absolute luminosity in the 0.2-12 keV range turns out to be 1.92 $\times 10^{34} (d(kpc)/10)^2$ erg s$^{-1}$ in the model with the powerlaw and 2.4 $\times 10^{34} (d(kpc)/10)^2$ erg s$^{-1}$
in the model with the partially covering absorber. 

An intriguing characteristics of two IP-novae previously observed in the years after the outburst was the supersoft portion of the spectrum, which indicated a very hot blackbody-type region present in novae V4743 Sgr and V2491 Cyg two years after the outburst \citep{Zemko2015, Zemko2016}, but not any more after 3.5 years in V4743 Sgr (V2491 Cyg was not observed again). The supersoft component was present in V407 Lup 3.5 years after the outburst;  although the nova distance is not well defined, we can say that the supersoft flux was approximately  an order of magnitude less luminous than in the other two novae. Moreover, this soft component can also be fitted with a thermal plasma with a temperature of 140-200 eV, instead of a blackbody at a temperature of $\simeq$100 eV, so it may not be due to residual nuclear burning. Without measuring emission lines we could not experiment with a detailed photoionization model, but it cannot be ruled out that the soft emission is due to photoionized gas near the cooling WD, perhaps already after turn-off of the burning. 

The spectral modeling leaves open possibilities, since we were able to fit the data with three different composite models. A fit can be obtained with two thermal components, of which one in the 150-200 eV range and the other peaking above the XMM-Newton range, with
 temperature of at least 19 keV. However, if we do not allow for a partially covering absorber with very high column density above 10$^{23}$ cm$^{-2}$ and covering fraction of 40\%-50\%, the maximum temperature becomes very high, reaching the maximum calculated value  of 68 keV in the XSPEC models. 
Moreover, the detection of several prominent lines in the low
 S/N RGS spectrum indicates that the soft portion of the spectrum also may be due to more than one component. The hotter thermal component is no longer necessary if we fit the spectrum with a flat powerlaw that produces most of the X-ray flux. 
We  suggest that this is actually an interesting aspect raised by the
{\sl XMM-Newton} observation, {\it  of an
 intriguing possibility, namely that  V407 Lup may be a
WD-pulsar}, like AR Sco and the more recently discovered AR Sco and eRASSU J191213.9-441044 \citep{Marsh2016, Pelisoli2024}, namely 
WDs with a strong magnetic field, in which particles
from the atmosphere of the close main sequence companion
can be accelerated.  Although a non-thermal component is not unambiguously detected in  WD-pulsars,
 they have
been discovered in close binaries (unlike ``regular'' neutron star pulsars that are often isolated) and the current interpretation of their
pulsations is that it is due to a beam of particles at the poles. 
At the distance of V407 Lup, discovering such peculiar objects is difficult, however
 a WD pulsar that is also actively accreting would facilitate the detection  of a peculiar non-thermal component thanks to additional X-ray flux due to accretion, and because of other phenomena (i.e. the nova outburst that has attracted attention). This
 is still highly speculative, but observations at higher energy, e.g. with {\sl NuSTAR}, would clarify this interesting issue.

The optical spectra confirm that a very hot photoionizing source was still present 5.5 years after the outburst. The equivalent width and the radial velocity of the H$\alpha$, H$\beta$ and  He II line at 4686 \AA  \  can be compared to the measurements during the late outburst by
\citet{Aydi2018}
The equivalent width of the He II line, between 4.7$\pm$0.5 and  9.7$\pm$0.8 \AA, is 
consistent with the low end of the single-component fits in 2017, although it seemed to have quite increased in 2018
\citep{Aydi2018}. The range of variation of the measured central wavelength corresponds to a velocity of about 50 km s$^{-1}$ for He II and about 100 km s$^{-1}$ for H$\alpha$, which may not be representative for the whole radial velocity range, but we note that it is quite smaller than the radial velocity measured in 2017/2018. 

The disappearance of the [O III] lines demonstrates that the nebular was dispersed significantly between 2019 and 2022, and probably this process was  already starting  2.5 years after the outburst, at the time of the last observation in 2019. There are very few optical spectra of novae repeated in time in the years after the outburst and this result is indicative of the velocity at which the remnant of the outflow
 dissipates in the interstellar medium, and of its density.

\section{Conclusions}
V407 Lup is one of a small, but growing group of IP-novae, so far comprising DQ Her \citep{Kemp1974}, GK Per \citep{Bianchini1983}, V4743 Sgr \citep{Leibowitz2006, Zemko2016, Zemko2018}, V2491 Cyg \citep{Zemko2015} and V1674 Her \cite{Drake2021}.
 The last three novae had eruptions in the era of X-ray satellites: they also showed distinct modulations  of the
SSS flux over the likely rotation period of the WD.

The atmosphere of the WD that is burning hydrogen in shell
 in a nova outburst cannot much hotter on the polar caps only because of accretion; the luminosity
 and the amplitude of the flux modulation during outburst in V407 Lup, like in V491 Cyg \citep{Ness2011}
 and V1674 Her \citep{Orio2022} clearly is not {\it not} due to an accretion column impinging the poles,  like we think it is the case for IPs in quiescence. It is
 possible that the burning itself occurs at a higher rate on the polar caps,
 or  that the atmospheric layer is less thick, and much
  hotter, at the bottom of the accretion columns. 
 Another intriguing common feature of these novae, described in the above papers, is that the supersoft component
 appeared to shrink before cooling, lasting for at least
 a few years with decreasing flux, but not with decreasing temperature. This
 is still another nova puzzle to solve, and it adds a layer of complexity to explore in the models:
 ``magnetic novae'' may account for several differences in outburst properties among novae. We will propose  new X-ray observations of V407 Lup and V2491 Cyg to find out whether the supersoft component faded completely like in nova V4743 Sgr, supporting the scenario of the shrinking, localized  burning region. Perhaps V407 Lup was observed exactly of the time of the final turn-off, and this is why the supersoft component is about an order of magnitude less intense than in V4743 Sgr and V2491 Cyg  2 years after their nova  outburst.

  The optical spectra show emission lines due to high ionization potential, with complex profiles that are quite typical of IPs. A study of the radial velocity and line profile variations over the optical period 
  was not possible with the SALT fixed altitude design, but it should  be another interesting and feasible project with a different telescope. 
\section*{Acknowledgements}
MO's study of the XMM data has been funded by the NASA-Goddard Space Flight center (award 80NSSC20K0827).
GJML is member of the CIC-CONICET (Argentina) and acknowledges support from grant ANPCYT-PICT 0901/2017
%%%%%%%%%%%%%%%%%%%%%%%%%%%%%%%%%%%%%%%%%%%%%%%%%%
\section*{Data Availability}
The X-ray data that support the findings of this study are 
 publicly available in the HEASARC and XMM-Newton web sites;
 the SALT data are available on request from the corresponding author.
%%%%%%%%%%%%%%%%%%%% REFERENCES %%%%%%%%%%%%%%%%%%

% The best way to enter references is to use BibTeX:

\bibliographystyle{mnras}
\bibliography{biblio} % if your bibtex file is called example.bib

% Alternatively you could enter them by hand, like this:
% This method is tedious and prone to error if you have lots of references
%\begin{thebibliography}{99}
%\bibitem[\protect\citeauthoryear{Author}{2012}]{Author2012}
%Author A.~N., 2013, Journal of Improbable Astronomy, 1, 1
%\bibitem[\protect\citeauthoryear{Others}{2013}]{Others2013}
%Others S., 2012, Journal of Interesting Stuff, 17, 198
%\end{thebibliography}

%%%%%%%%%%%%%%%%%%%%%%%%%%%%%%%%%%%%%%%%%%%%%%%%%%

%%%%%%%%%%%%%%%%% APPENDICES %%%%%%%%%%%%%%%%%%%%%

\appendix

%%%%%%%%%%%%%%%%%%%%%%%%%%%%%%%%%%%%%%%%%%%%%%%%%%

% Don't change these lines
\bsp	% typesetting comment
\label{lastpage}
\end{document}